\documentclass{PoS}

\input{standard_preamble}

\rightline{\parbox{4cm}{\large\rm
    ADP-15-42/T944 \\
    DESY 15-214 \\
    Edinburgh 2015/28 \\
    LTH 1067
  }}

\title{Applications of the Feynman-Hellmann theorem in hadron structure}

\ShortTitle{Feynman-Hellmann \& hadron structure}

\author{
  A.~J.~Chambers,$^a$
  R.~Horsley,$^b$
  Y.~Nakamura,$^c$
  H.~Perlt,$^d$
  D.~Pleiter,$^{ef}$
  P.~E.~L.~Rakow,$^g$
  G.~Schierholz,$^h$
  A.~Schiller,$^d$
  H.~St\"uben,$^i$
  \speaker{R.~D.~Young}$^{aj}$
  and
  J.~M.~Zanotti$^a$ \\
  \llap{$^a$}
  CSSM, Department of Physics, University of Adelaide, Adelaide SA
  5005, Australia \\
  \llap{$^b$}
  School of Physics and Astronomy, University of Edinburgh, Edinburgh
  EH9 3JZ, UK\\
  \llap{$^c$}
  RIKEN Advanced Institute for Computational Science, Kobe, Hyogo
  650-0047, Japan\\
  \llap{$^d$}
  Institut f\"ur Theoretische Physik, Universit\"at Leipzig, 04103
  Leipzig, Germany\\
  \llap{$^e$}
  JSC, J\"ulich Research Centre, 52425 J\"ulich, Germany\\
  \llap{$^f$}
  Institut f\"ur Theoretische Physik, Universit\"at Regensburg, 93040
  Regensburg, Germany\\
  \llap{$^g$}
  Theoretical Physics Division, Department of Mathematical Sciences
  University of Liverpool, Liverpool L69 3BX, UK\\
  \llap{$^h$}
  Deutsches Elektronen-Synchrotron DESY, 22603 Hamburg, Germany\\
  \llap{$^i$}
  Regionales Rechenzentrum, Universit\"at Hamburg, 20146 Hamburg,
  Germany\\
  \llap{$^j$}
  CoEPP, Department of Physics, University of Adelaide, Adelaide SA
  5005, Australia \\
}

\abstract{
  The Feynman-Hellmann (FH) relation offers an alternative way
  of accessing hadronic matrix elements through artificial modifications
  to the QCD Lagrangian.
  In particular, a FH-motivated method provides a new approach to
  calculations of disconnected contributions to matrix elements and
  high-momentum nucleon and pion form factors.
  Here we present results for the total nucleon axial charge,
  including a statistically significant non-negative total
  disconnected quark
  contribution of around $-5\%$ at an unphysically heavy pion mass.
  Extending the FH relation to finite-momentum transfers,
  we also present calculations of the pion and nucleon electromagnetic
  form factors up to momentum transfers of around 7--8
  GeV\textsuperscript{2}.
  Results for the nucleon are not able to confirm the existence of a
  sign change for the ratio $\frac{G_E}{G_M}$, but suggest that future
  calculations at lighter pion masses will provide fascinating insight
  into this behaviour at large momentum transfers.
}

\FullConference{The 33rd International Symposium on Lattice Field Theory\\
		14 -18 July 2015\\
		Kobe International Conference Center, Kobe, Japan}

\begin{document}

\section{Introduction}

In recent years, lattice calculations have seen much success in
reproducing the hadron spectrum, with increasing levels of precision.
However with regards to investigations of hadron structure, there are
still many outstanding issues that have yet to be fully dealt with.
In particular, open questions include the details of the origin
of hadronic spin and the
short-distance distribution of charge within hadrons.

In 1987, results obtained by the European
Muon Collaboration (EMC) \cite{Ashman:1987hv} (and more recently
COMPASS \cite{Alexakhin:2006oza}) showed that the quark
spin degrees of freedom carry only a small fraction of the overall
nucleon spin.
Since then, a huge effort has been expended to reproduce this result
theoretically, with lattice playing an important role.
One challenge in lattice calculations of the nucleon axial matrix
elements is the difficulty of accessing fermion line disconnected
contributions.
There has been significant progress in this area, with the use
of stochastic methods to attempt calculations of these
quantities \cite{Babich:2010at,QCDSF:2011aa,Engelhardt:2012gd,
  Abdel-Rehim:2013wlz,Deka:2013zha} (also \cite{Green:2015wqa}
for the vector matrix element). However in addition to this, there is still much ongoing
discussion regarding the need to
control potential excited-state contamination effects in three-point
function calculations of the connected contributions
\cite{Owen:2012ts,Capitani:2012gj,Dinter:2011sg,Bhattacharya:2013ehc}.

Experimental determinations of high-momentum form factors for the
nucleon and pion are difficult due to the lack of free neutron or pion
targets. The Rosenbluth separation method
also has trouble isolating the electric form factor of the proton at high momenta.
In lattice, it is difficult to
extract clean signals for the vector matrix element
from the noise introduced by large momentum projections.
Additionally, lattice studies must address the same issues of
excited-state contamination control and disconnected contributions
as in the axial case.

Recently we have carried out work using a Feynman-Hellmann (FH)
motivated method to address some of these issues.
This approach uses
the FH theorem to calculate hadronic matrix elements in lattice QCD
through modifications to the QCD Lagrangian.
The method has already seen great success
in calculations of connected contributions to axial matrix
elements \cite{Chambers:2014qaa},
gluon observables \cite{Horsley:2012pz},
and singlet renormalisation factors \cite{Chambers:2014pea}.
Here we show recent determinations of the disconnected
contributions to nucleon spin
(discussed in detail in \cite{Chambers:2015disconn}), and
present preliminary new results for the
nucleon and pion electromagnetic form factors.

\section{Feynman-Hellmann Methods}

The FH method relates shifts in the hadron spectrum,
as an external field is applied, to hadronic matrix elements.
In this way, it
allows quantities traditionally accessed using 3-point functions to be
calculated from 2-point functions alone.
With an additional operator in the QCD Lagrangian,
\begin{equation}
  \Lagrangian \to
  \Lagrangian + \lambda \mathcal{O}
  \eqcomma
  \label{eq:general_lag_mod}
\end{equation}
we have
(up to choices of correlator projectors for states with
non-trivial Dirac structure),
\begin{equation}
  \left. \frac{\partial E}{\partial \lambda} \right|_{\lambda = 0}
  =
  \frac{1}{2E} \bra{ H } \mathcal{O} (0) \ket{H}
  \eqcomma
\end{equation}
for any hadron state $H$.
Hence we see that on the lattice, a particular hadronic matrix element may be
calculated by performing hadron spectroscopy for multiple values of
$\lambda$, and observing the linear behaviour in the resulting energy
shifts about $\lambda=0$.

The modification in \eq{general_lag_mod} may be made either during gauge
field generation, or during the inversion of the Dirac operator
matrix.
The first case necessitates the generation of new field ensembles for
several values of $\lambda$, for each operator,
and accesses fermion line disconnected contributions to the matrix element.
The second case makes use of existing ensembles, and accesses fermion line
connected contributions.

\section{Simulation Details}

For this work we use gauge configurations with $2+1$ flavours of
non-perturbatively $O(a)$-improved Wilson fermions, where
the lattice spacing $a \approx 0.074$ fm is set using a number of singlet
quantities \cite{Bornyakov:2015eaa,Bietenholz:2010jr,Bietenholz:2011qq}.
The clover action comprises the tree-level Symanzik improved
gluon action together with a stout smeared fermion action, modified
for the use of the FH technique
\cite{Chambers:2014qaa}.

For the disconnected results in \sec{disconn} we use ensembles with two sets
of hopping parameters,
$(\kappa_l,\kappa_s) = (0.120900, 0.120900) \text{ and }
(0.121095,0.120512)$, the first of these corresponding to the SU(3)
flavour symmetric point.
These both have a lattice volume of $L^3 \times T = 32^3 \times 64$,
and the corresponding pion masses are approximately 470 and 310 MeV
respectively (in the $\lambda=0$ limit).
The ensembles are
generated with the modified quark action described in
\eq{disconn_mod}.
For further details of these ensembles, and the simulated values of
$\lambda$, see \cite{Chambers:2015disconn}.

Two volumes
($L^3 \times T = 24^3 \times 48
\text{ and } 32^3 \times 64$)
are used for the form factor calculations in \sec{form_factors},
both with hopping parameters $(\kappa_l, \kappa_s) = (0.120900, 0.120900)$
corresponding to the SU(3) flavour symmetric point.
The ensembles have been generated with an unmodified quark action,
with the Dirac matrix modified during propagator calculation instead.
We use two different values of $\lambda$, and several different
sets of kinematics.

\section{Disconnected Calculations}
\label{sec:disconn}

The FH approach offers an alternative technique for
accessing disconnected contributions
to matrix elements.
For an analysis of disconnected contributions to the axial operator,
we modify the fermion part of the QCD
Lagrangian during gauge field generation such that
\begin{equation}
  \Lagrangian \to
  \Lagrangian + \lambda \sum_{q=u,d,s} \bar{q} \gamma_3 \gamma_5 q
  \eqstop
  \label{eq:disconn_mod}
\end{equation}
Note that all three simulated flavours are modified
simultaneously.
% This is motivated by the expectation
% that the signal-to-noise ratio of the larger combined signal
% is better than that of the individual flavour contributions in isolation.
It is expected that the larger combined signal will give a better
signal-to-noise ratio than the individual flavour contributions.

\begin{figure}
  \centering
  \begin{minipage}{0.45\textwidth}
    \centering
    \includegraphics[width=\columnwidth]{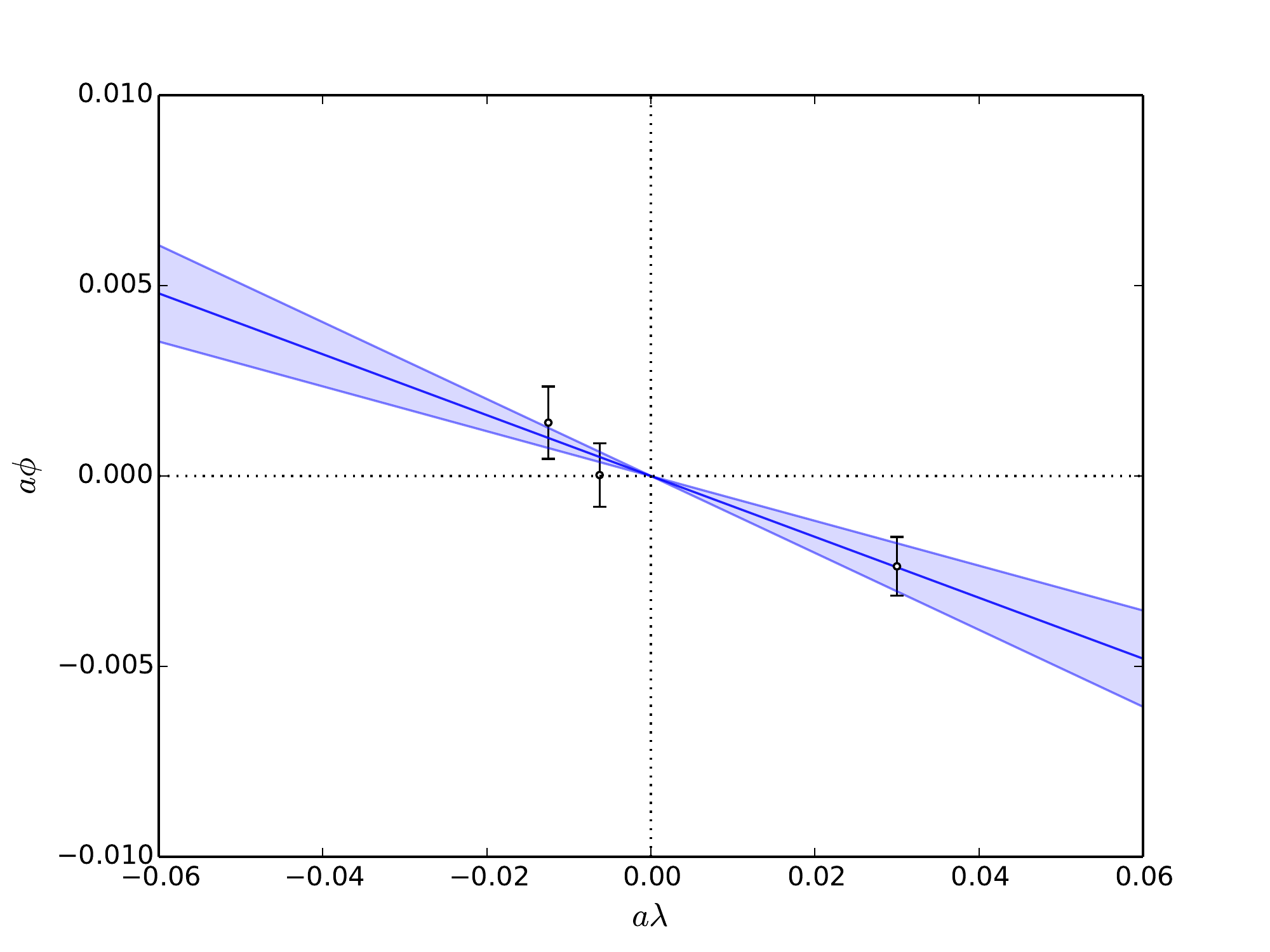}
    \caption{Correlator phase on each ensemble as a function of
      $\lambda$, with the axial operator term included in the fermion
      action during gauge field generation with
      $(\kappa_l, \kappa_s) = (0.120900, 0.120900)$.
      The slope is proportional to
      the quantity $\Delta \Sigma_\text{disc.}$}
    \label{fig:disconn_sym}
  \end{minipage}\hfill
  \begin{minipage}{0.45\textwidth}
    \centering
    \includegraphics[width=\columnwidth]{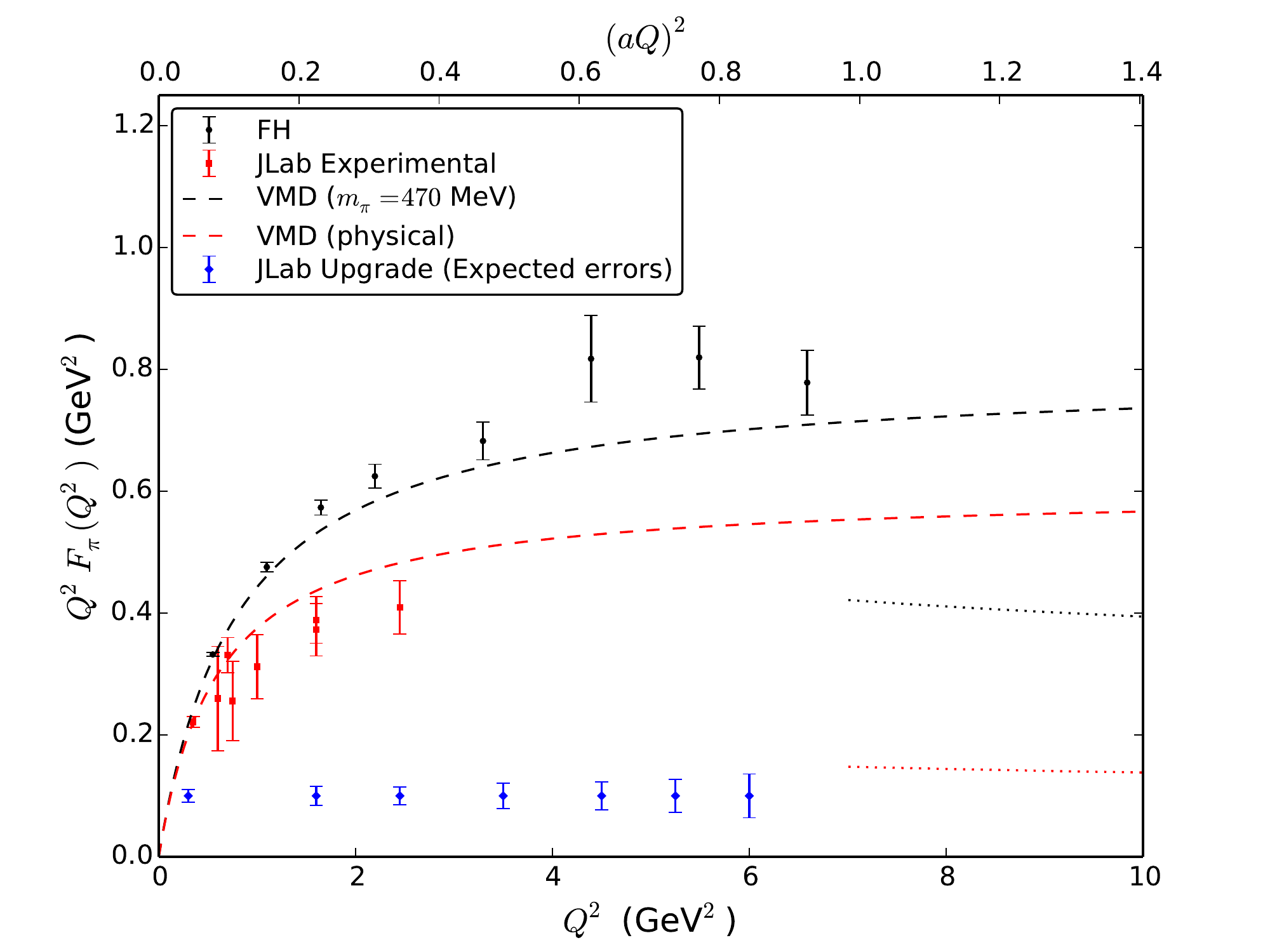}
    \caption{Pion form factor as a function of momentum transfer from
      both the FH method and JLab experimental data. The
      vector meson dominance is included to guide the eye.}
    \label{fig:q2fpi}
  \end{minipage}
\end{figure}

With the additional operator in \eq{general_lag_mod}, we expect the
FH signal to appear in the imaginary part of the nucleon
correlation function.
Introducing an additional imaginary factor in
\eq{disconn_mod} would shift this signal into the real channel,
however the resulting operator is not $\gamma_5$-Hermitian, and hence
introduces a sign problem.
The nucleon correlator on the modified ensembles
takes the form (to first order in $\lambda$),
\begin{equation}
  G_\pm(\lambda, t)
  \atlarget
  \left[
    A \pm \lambda(\Delta A + i\Delta B)
  \right]
  e^{- \left[ E \pm i \lambda \Delta \Sigma_\text{disc.} \right] t}
  \eqcomma
  \label{eq:disconn_corr}
\end{equation}
where the disconnected contribution to the total quark spin fraction
\begin{equation}
  \Delta \Sigma_\text{disc.} =
  \Delta u_\text{disc.} + \Delta d_\text{disc.} + \Delta s \eqcomma
\end{equation}
and $\Delta A$ and $\Delta B$ are arbitrary real numbers parameterising
the shift in the correlation amplitude.
The $\pm$ signs correspond to spin-up and down positive parity projections
of the lattice nucleon state respectively.
To extract $\Delta \Sigma_\text{disc.}$, we form a ratio of
correlators on each ensemble,
which is approximately linear in $t$
(to first order in $\lambda$)
for large lattice times,
\begin{equation}
  R(\lambda, t)
  =
  \frac
  {\Im \left[ G_-(\lambda,t) - G_+(\lambda,t) \right] }
  {\Re \left[ G_-(\lambda,t) + G_+(\lambda,t) \right] }
  \atlarget
  \lambda \Delta \Sigma_\text{disc.} t
  - \lambda \frac{\Delta B}{A}
  \eqstop
  \label{eq:disconn_ratio}
\end{equation}
We introduce an effective phase shift, which at large times
indicates ground state saturation,
\begin{equation}
  \phi_{\text{eff.}} = \frac{1}{a} \left[ R(\lambda, t+a) - R(\lambda,
    t) \right]
  \atlarget
  \lambda \Delta \Sigma_\text{disc.} \eqstop
\end{equation}
\fig{disconn_sym} shows the change in the correlator phase with
changing $\lambda$ at the SU(3) flavour symmetric point,
extracted using the described methods.
By fitting the linear dependence
of the phase shift,
and repeating the analysis at the lighter mass,
we calculate the quark disconnected
contribution to $\Delta \Sigma$ (renormalised according to
determinations of the singlet and non-singlet factors in
\cite{Chambers:2014pea}) of
\begin{align}
  \Delta \Sigma^{\overline{\text{MS}}(2 \text{ GeV})}_\text{disc.} \;
  (m_\pi \approx 470 \text{ MeV}) & = -0.055(18) \eqcomma \\
  \Delta \Sigma^{\overline{\text{MS}}(2 \text{ GeV})}_\text{disc.} \;
  (m_\pi \approx 310 \text{ MeV}) & = \phantom{-}0.026(14) \eqstop
\end{align}
The result at the heavier mass is consistent with existing results,
however the unusual behaviour at the lighter mass remains to be quantified.

\section{Form Factors}
\label{sec:form_factors}

By including a non-zero momentum injection in the operator insertion,
we are able to apply the FH approach to
calculate more general non-forward matrix elements.
We make the modification to the Lagrangian density
\begin{equation}
  \Lagrangian(x) \to
  \Lagrangian(x) + \lambda e^{i \vec{q} \cdot (\vec{x} - \vec{x_0})}
  \bar{q}(x) \gamma_\mu q(x)
  \eqcomma
  \label{eq:form_factor_lag_mod}
\end{equation}
where the modification is made to the $u$ and $d$ quarks separately.
This change is only applied to the Dirac fermion operator before
inversion,
and hence we only access connected contributions to the vector matrix element.
Note that the momentum insertion is made with respect to some spatial location
$\vec{x}_0$, and hence we restrict the space source location for quark
propagators to this point.
Our application of the FH relation
requires momentum projection at the propagator sink
to enforce Breit frame kinematics.
This choice acts to avoid an energy transfer through the current,
which would lead to a non-trivial
time dependence of the correlation function.

We perform the operator insertion in \eq{form_factor_lag_mod}
for two components of the vector
current ($\mu = 2,4$), and several different momentum transfers.
For the majority of the chosen kinematics, the total
3-vector momentum is zero, with $\vecp{p} = - \vec{p}$.
It is this simple case that is discussed here,
with more complicated momentum transfers to be discussed in a future publication.
These kinematics, in addition to the Breit frame requirement,
allow us to minimise momentum projection at the propagator sink, which
appears to reduce the noise introduced to the correlation function.

\subsection{Pion}

% The pion vector matrix element is parameterised by a single form
% factor $F_\pi$,
% the individual quark contributions being given by
Individual quark contributions to the pion vector form factor $F_\pi$ are
given by
\begin{equation}
  \bra{ \pi(\vecp{p}) }
  \bar{q}(0) \gamma_\mu q(0)
  \ket{ \pi(\vec{p}) }
  =
  (p_\mu + p_\mu') F_{\pi q}(Q^2)
  \eqstop
\end{equation}
Note that we work in the isospin symmetric limit, and consider only the
modification to the $u$ quark in the $\pi^+$ state. The $d$ quark
result comes from charge conjugation.
With the modification to the Lagrangian in \eq{form_factor_lag_mod},
considering only the temporal component of the vector current, we have
\begin{equation}
  \left. \frac{\partial E}{\partial \lambda} \right|_{\lambda = 0}
  =
  F_\pi(Q^2)
  \eqstop
\end{equation}
For the spatial component of the vector current, the first order shift
in the energy vanishes when $\vecp{p} + \vec{p} = \vec{0}$, and so we
only use results from the temporal current insertion.

\fig{q2fpi} shows results obtained for $Q^2F_\pi(Q^2)$ at a
variety of momentum transfers, compared with experimental data from
JLab.
The form factor is anticipated to be quite sensitive to the
quark mass, however the overall trend is in good agreement with naive vector
meson dominance.
A discussion of the asymptotic transition in the context of \cite{Chang:2013nia}
will feature in an upcoming publication.
Included on the plot are the $Q^2$ targets for the JLab upgrade,
and expected error ranges.
We note that lattice results at such large momenta
may be particularly complimentary to such experiments.

\subsection{Nucleon}

Individual quark contributions to the Dirac and Pauli form factors
($F_1$ and $F_2$ respectively) for the nucleon are defined by the matrix elements
\begin{equation}
  \bra{N(\vecp{p}, \vecp{s})}
  \bar{q}(0) \gamma_\mu q(0)
  \ket{N(\vec{p}, \vec{s})}
  =
  \bar{u}(\vecp{p}, \vecp{s})
  \left[
    \gamma_\mu F_{1q}(Q^2)
    + \sigma_{\mu \nu} \frac{q_\nu}{2 m} F_{2q}(Q^2)
  \right]
  u(\vec{p}, \vec{s})
  \eqstop
\end{equation}
For the temporal component of the current insertion we
choose projection matrices to project unpolarised positive
parity states,
and for the spatial component we project spin up and down positive
parity states individually.
The resulting energy shifts from the FH relation are in general linear
combinations of $F_1$ and $F_2$, and the results from both current
insertions can be combined to extract the two form factors
individually.
However for the case $\vecp{p} = - \vec{p}$,
the FH analysis is greatly simplified.
The shifts from the insertion of the temporal and spatial currents are
directly proportional to the Sachs electric and magnetic form factors respectively,
\begin{equation}
  \left. \frac{\partial E}{\partial \lambda} \right|_{\lambda = 0}
  =
  \frac{m}{E} G_{Eq}(Q^2)
  \qquad \quad \text{and} \quad \qquad
  \left. \frac{\partial E}{\partial \lambda} \right|_{\lambda = 0} =
  \pm \frac{q_1}{2E} G_M(Q^2)
  \eqstop
\end{equation}
Here $q_1$ appears in the spatial case as a result of choosing the second
spatial component of the vector current, and the $z$-axis
as the spin quantisation axis.

\begin{figure}
  \centering
  \includegraphics[width=0.49\columnwidth]{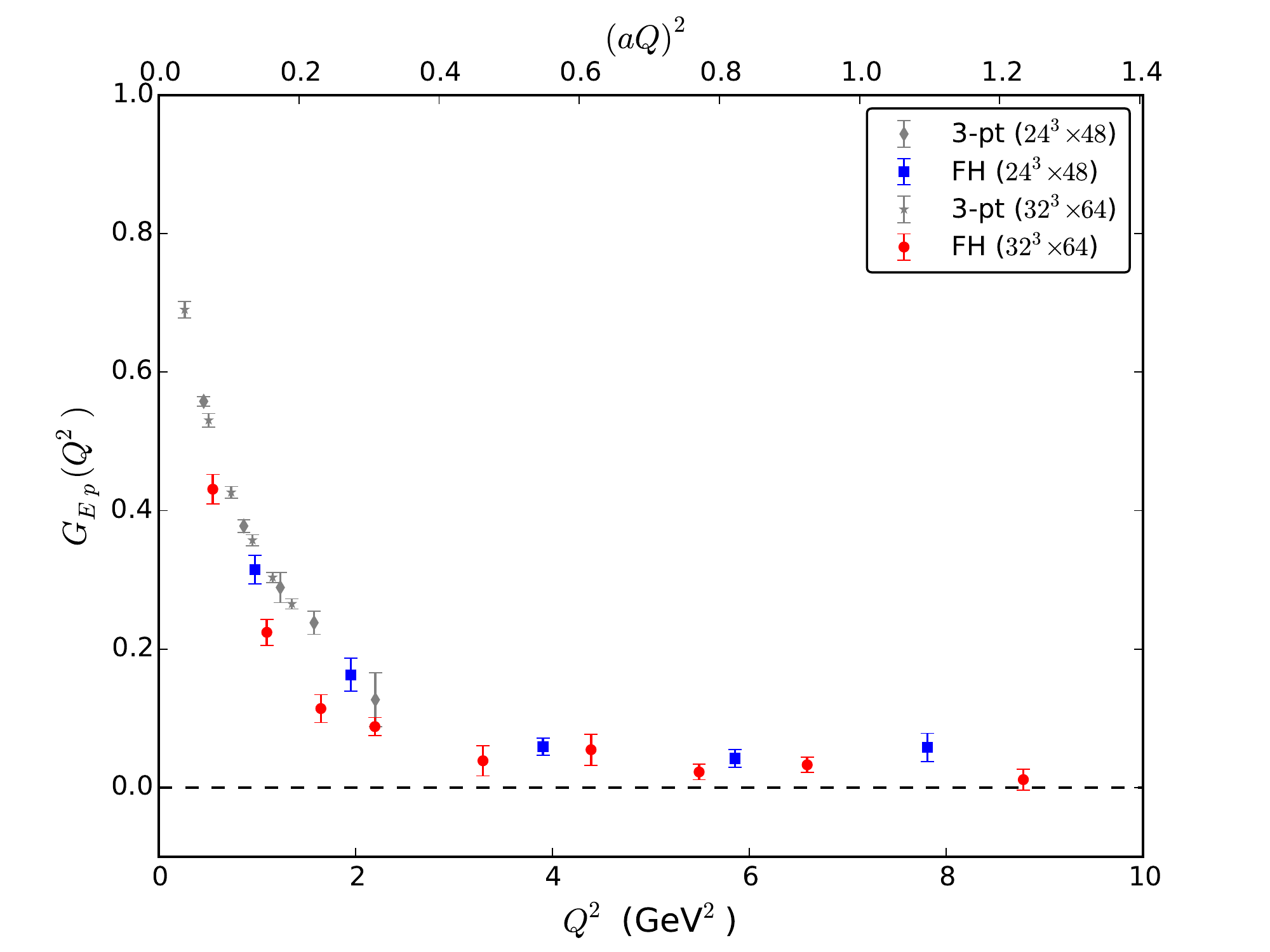}
  \includegraphics[width=0.49\columnwidth]{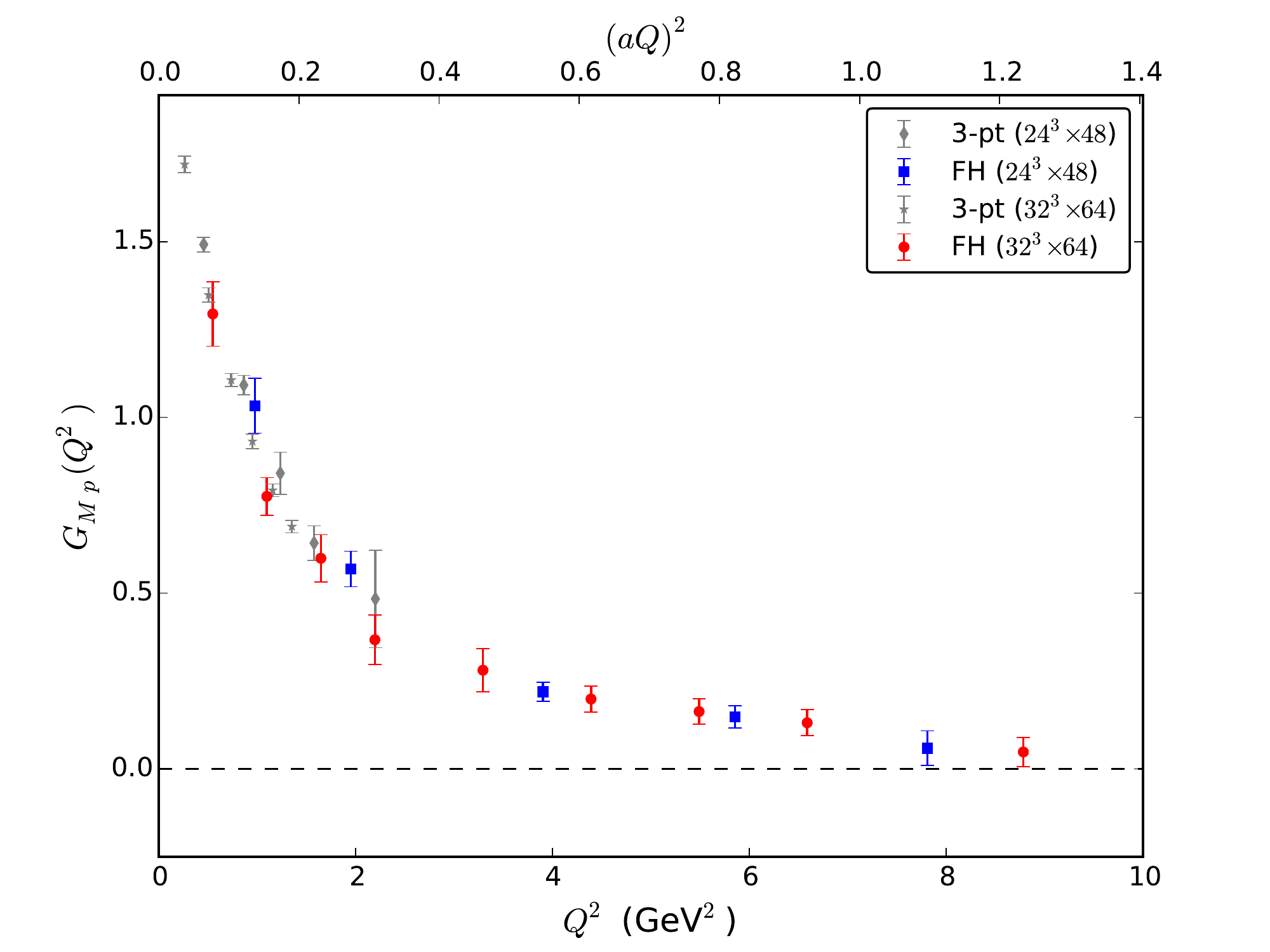}
  \caption{Electric and magnetic form factors for the proton,
    calculated using the FH method and a standard
    3-point function method, on two different lattice sizes.}
  \label{fig:ge_and_gm}
\end{figure}

\fig{ge_and_gm} shows results for the electric and magnetic form
factors of the proton, determined by combining results from the
individual $u$ and $d$ contributions.
We note very good agreement at
low $Q^2$ between the standard 3-point function method and the FH
approach,
and particularly draw attention to the very promising
results at high $Q^2$. Although the current precision is not sufficient to
confirm or deny the existence of a zero crossing in the ratio $G_{Ep}/G_{Mp}$,
we are optimistic about the possibilities of higher statistics
calculations at lighter quark masses.

\section{Summary}

We have shown how a FH approach can be used to perform a
full high-precision calculation of the axial matrix element of the
nucleon.
We have also shown how the method may be extended to the calculation
of non-forward hadron matrix elements.
The procedures described may easily be applied to other hadron
observables.

Further extensions of the nucleon and pion form factor
calculations will attempt to investigate even higher momenta than
those already accessed, in addition to explorations of the effect of
lattice systematics on such simulations. These will be important for
contributing understanding to ongoing experimental efforts to access
such scales.

\acknowledgments

The numerical configuration generation was performed using the BQCD
lattice QCD program, \cite{Nakamura:2010qh}, on the IBM BlueGeneQ
using DIRAC 2 resources (EPCC, Edinburgh, UK), the BlueGene P and Q at
NIC (J\"ulich, Germany) and the Cray XC30 at HLRN
(Berlin--Hannover, Germany).
Some of the simulations were undertaken using resources awarded at the
NCI National Facility in Canberra, Australia, and the iVEC facilities
at the Pawsey Supercomputing Centre. These resources are provided
through the National Computational Merit Allocation Scheme and the
University of Adelaide Partner Share supported by the Australian
Government.
The BlueGene codes were optimised using Bagel \cite{Boyle:2009vp}.
The Chroma software library \cite{Edwards:2004sx}, was used in the
data analysis.
This investigation has been supported
by the Australian Research Council under grants
FT120100821, FT100100005 and DP140103067.
HP was supported by DFG grant SCHI 422/10-1.

% \begin{thebibliography}{99}
% \bibitem{...}
% ....

% \end{thebibliography}

\bibliography{ref}

\begin{thebibliography}{10}

\bibitem{Ashman:1987hv}
J.~Ashman {\em et~al.},
\newblock Phys.~Lett. {\bf B206}, 364 (1988).

\bibitem{Alexakhin:2006oza}
V.~{\relax Yu}. Alexakhin {\em et~al.},
\newblock Phys.~Lett. {\bf B647}, 8 (2007), hep-ex/0609038.

\bibitem{Babich:2010at}
R.~Babich {\em et~al.},
\newblock Phys.~Rev. {\bf D85}, 054510 (2012), 1012.0562.

\bibitem{QCDSF:2011aa}
G.~S. Bali {\em et~al.},
\newblock Phys.~Rev.~Lett. {\bf 108}, 222001 (2012), 1112.3354.

\bibitem{Engelhardt:2012gd}
M.~Engelhardt,
\newblock Phys.~Rev. {\bf D86}, 114510 (2012), 1210.0025.

\bibitem{Abdel-Rehim:2013wlz}
A.~Abdel-Rehim {\em et~al.},
\newblock Phys.~Rev. {\bf D89}, 034501 (2014), 1310.6339.

\bibitem{Deka:2013zha}
M.~Deka {\em et~al.},
\newblock Phys. Rev. {\bf D91}, 014505 (2015), 1312.4816.

\bibitem{Green:2015wqa}
J.~Green {\em et~al.},
\newblock Phys. Rev. {\bf D92}, 031501 (2015), 1505.01803.

\bibitem{Owen:2012ts}
B.~J. Owen {\em et~al.},
\newblock Phys.~Lett. {\bf B723}, 217 (2013), 1212.4668.

\bibitem{Capitani:2012gj}
S.~Capitani {\em et~al.},
\newblock Phys.~Rev. {\bf D86}, 074502 (2012), 1205.0180.

\bibitem{Dinter:2011sg}
S.~Dinter {\em et~al.},
\newblock Phys.~Lett. {\bf B704}, 89 (2011), 1108.1076.

\bibitem{Bhattacharya:2013ehc}
T.~Bhattacharya {\em et~al.},
\newblock Phys. Rev. {\bf D89}, 094502 (2014), 1306.5435.

\bibitem{Chambers:2014qaa}
A.~J. Chambers {\em et~al.},
\newblock Phys.~Rev. {\bf D90}, 014510 (2014), 1405.3019.

\bibitem{Horsley:2012pz}
R.~Horsley {\em et~al.},
\newblock Phys.~Lett. {\bf B714}, 312 (2012), 1205.6410.

\bibitem{Chambers:2014pea}
A.~J. Chambers {\em et~al.},
\newblock Phys.~Lett. {\bf B740}, 30 (2015), 1410.3078.

\bibitem{Chambers:2015disconn}
A.~J. Chambers {\em et~al.},
\newblock (2015), 1508.06856.

\bibitem{Bornyakov:2015eaa}
V.~G. Bornyakov {\em et~al.},
\newblock (2015), 1508.05916.

\bibitem{Bietenholz:2010jr}
W.~Bietenholz {\em et~al.},
\newblock Phys.~Lett. {\bf B690}, 436 (2010), 1003.1114.

\bibitem{Bietenholz:2011qq}
W.~Bietenholz {\em et~al.},
\newblock Phys.~Rev. {\bf D84}, 054509 (2011), 1102.5300.

\bibitem{Chang:2013nia}
L.~Chang {\em et~al.},
\newblock Phys.~Rev.~Lett. {\bf 111}, 141802 (2013), 1307.0026.

\bibitem{Nakamura:2010qh}
Y.~Nakamura and H.~St{\"u}ben,
\newblock PoS {\bf LATTICE2010}, 040 (2010), 1011.0199.

\bibitem{Boyle:2009vp}
P.~A. Boyle,
\newblock Comput.~Phys.~Commun. {\bf 180}, 2739 (2009).

\bibitem{Edwards:2004sx}
R.~G. Edwards and B.~Joo,
\newblock Nucl.~Phys.~Proc.~Suppl. {\bf 140}, 832 (2005), hep-lat/0409003.

\end{thebibliography}

\end{document}